\documentclass[floatfix,%
 reprint,
 amsmath,amssymb,
 aps,
]{revtex4-2}

\usepackage{graphicx}% Include figure files
\usepackage{dcolumn}% Align table columns on decimal point
\usepackage{bm}% bold math
\usepackage{color}
\usepackage{color}
\usepackage{siunitx}
\usepackage{hyperref}% add hypertext capabilities
\begin{document}

% Published version — DOI added for arXiv update
\title{Nonlinear analysis of causality for heat flow in heavy-ion collisions: constraints from equation of state}% Force line breaks with \\
\thanks{Published in Physical Review; DOI:\,\href{https://link.aps.org/doi/10.1103/ftyb-dps7}{10.1103/ftyb-dps7}}%
\author{Victor Roy}
\affiliation{School of Physical Sciences, National Institute of Science Education and Research, Bhubaneswar 752050, India.}
\affiliation{Homi Bhabha National Institute, Training School Complex, Anushaktinagar, Mumbai 400094, Maharashtra, India.}

\date{\today}% It is always \today, today,

\begin{abstract}
The present work investigates the causal parameter space of the Mueller-Israel-Stewart second-order theory for  heat-conducting fluids in the Eckart frame for one-dimensional fluid
flow in systems with finite baryon density.
It is shown that this parameter space is highly constrained and particularly sensitive to the equation of state and second-order transport coefficients. Through numerical analysis of the characteristic equations, the present analysis identifies regions of strong hyperbolicity, weak hyperbolicity, and non-hyperbolicity, mapping the boundaries of causality violation as functions of the heat flux to energy density ratio $q/\varepsilon$ and relaxation parameters. The present work also explores the causality conditions using a realistic lattice QCD-based equation of state. Using the Navier-Stokes approximation, an estimate is made of the heat flow magnitude to assess causality criteria for one-dimensional heat conduction in heavy-ion collisions. The present calculations reveal unrealistically large heat flux values ($|{\bf{q}}|/\varepsilon \approx 330$--$811$) for typical RHIC conditions when using thermal conductivity estimates from kinetic theory models, suggesting either significant overestimation of transport coefficients or breakdown of the fluid approximation in these extreme conditions. The pressure gradient corrections reduce the heat flow by approximately 15\% but do not resolve the causality concerns.
\end{abstract}

\maketitle

%\tableofcontents

\section{Introduction}

% The study of relativistic dissipative fluid dynamics has gained significant attention due to its relevance in describing extreme physical conditions such as those in neutron stars and ultrarelativistic heavy-ion collisions. Among the most established frameworks for relativistic dissipation, the oldest and most widely used is the Müller-Israel-Stewart (MIS) theory \cite{MIS1,MIS2}, which addresses the acausality and instability issues inherent in the straightforward relativistic generalization of the Navier-Stokes equations \cite{NS1,NS2}. Recent improved formulations include the DNMR theory \cite{Denicol2012}, developed by Denicol, Niemi, Molnár, and Rischke; the BRSSS theory developed by Baier, Romatschke, Son, Starinets, and Stephanov \cite{Baier2008}; and the BDNK framework, introduced by Bemfica, Disconzi, Noronha, and Kovtun \cite{Bemfica2019}.

The study of relativistic dissipative fluid dynamics has gained significant attention due to its relevance in describing extreme physical conditions such as those in neutron stars and ultra-relativistic heavy-ion collisions. Among the most established frameworks for relativistic dissipation, the oldest and most widely used is the Mueller-Israel-Stewart (MIS) theory \cite{MIS1,MIS2}, which addresses the acausality and instability issues inherent in the straightforward relativistic generalization of the Navier-Stokes equations \cite{NS1,Eckart1940}. Among some of the recent improved formulations are the DNMR theory \cite{Denicol2012}, developed by Denicol, Niemi, Molnar, and Rischke; the BRSSS theory developed by Baier, Romatschke, Son, Starinets, and Stephanov \cite{Baier2008}; and the BDNK framework, introduced by Bemfica, Disconzi, Noronha, and Kovtun \cite{Bemfica2019}.

DNMR is a second-order relativistic hydrodynamics formulation derived from kinetic theory using the 14-moment approximation. It extends the Mueller-Israel-Stewart (MIS) framework by truncating the moment equations to include only the slowest eigenmodes of the linearized collision integral, reducing the number of dynamical variables while maintaining causality and stability \cite{Denicol2012}. BRSSS is also a second-order relativistic hydrodynamics formulation that ensures causality and conformal invariance, particularly in strongly coupled systems like the quark-gluon plasma (QGP). Meanwhile, BDNK is a first-order approach that still offers causality and stability \cite{Pandya2022,Bemfica2022}.

MIS theory, rooted in extended irreversible thermodynamics \cite{EIT}, introduces relaxation times for dissipative currents, ensuring a causal and stable evolution under specific conditions \cite{Baier2008} and is one of the most widely used formulations of dissipative
hydrodynamics in heavy-ion collisions.

%%%%%%%%%%%%%%%%%%%%%%%%%%%%%%
The importance of causality in relativistic hydrodynamics originates from the requirement that signal propagation must not exceed the speed of light. This condition is mathematically reflected in the hyperbolicity of the equations of motion, which is guaranteed when the eigenvalues of the characteristic determinant are real, positive and distinct \cite{RezzollaZanotti2013}. Although many previous studies~\cite{SusskindCausality,KrotscheckKundt1978,Gavassino2020,Gavassino:2025bxx,Grossi2018,Pu2010,Biswas:2022hiv,Biswas:2020rps,Romatschke2010,Sammet2023,Bhattacharyya:2024tfj,Spalinski2023,Perna2021,Brito2022,BDN2019,Wang2024,Hiscock1985,Hoshino:2024qun} have analyzed causality and stability mostly in the linear regime, including the effects of shear stress and net charge diffusion, there remains significant scope to explore the hyperbolicity and causality of MIS theory in the context of heat conducting fluids, particularly in the non-linear regime. Recent extensions have begun addressing nonlinear causality in contexts including diffusion, magnetohydrodynamics, and constraints on initial conditions for one-dimensional expanding fluids~\cite{Cordeiro:2025mtg,CordeiroPRL2024,Hoshino:2024qun}.

This issue is particularly relevant for systems exhibiting significant heat flux \cite{Gavassino2024}. As noted by Hiscock and Lindblom \cite{Hiscock1988}, the strict hyperbolicity of the equations of motion can break down when the heat flux becomes extremely large, with the ratio $|q|/\varepsilon$ reaching some critical values. For instance, they reported that for $|q|/\varepsilon \approx 0.08898$ for a conformal fluid, the fluid equations cease to be hyperbolic, corresponding to a regime of extraordinarily large heat flux unlikely to be encountered in realistic physical systems. Additionally, they showed that varying transport coefficients, such as adopting a ten times larger relaxation time through parameter \(\beta_1\), can extend the hyperbolicity domain to $|q|/\varepsilon \approx 1/3$.

In ultrarelativistic heavy-ion collisions, where deviations from equilibrium can be substantial, it is important to explore whether the heat flux values and causality limits predicted by the MIS framework fall within physically realistic ranges. One of the aims of this work is to estimate the typical order of magnitude of heat flux in such systems and investigate the extent of the causality-constrained parameter space.
It is shown that with the current estimate of the coefficient of thermal conductivity the heat flow could reach extremely large values for central fireball temperatures as low as 150 MeV. It is well known that the choice of hydrodynamic frame for defining the fluid four-velocity in dissipative fluids can lead to specific cases such as the Landau frame, where heat flow does not contribute to the energy flux $T^{0i}$ but reappears as a finite baryon current that is proportional to the thermal conductivity and the gradient of the ratio of chemical potential to temperature~\cite{PhysRevD.31.53}. Conversely, one can choose a different frame, such as the Eckart frame, where the dissipative part of the baryon four-current vanishes and the energy flux becomes non-vanishing~\cite{Eckart1940,KunihiroTsumura2008,Monnai2019}.

In this work, the hyperbolicity of the MIS second-order theory for a heat-conducting fluid in Eckart frame is investigated. The present analysis focuses on the dependence of the causality space on the equation of state (EoS), which relates the thermodynamic variables of the fluid. Specifically, the characteristic propagation speed for a one dimensional fluid flow is examined as a function of the ratio of heat flux to fluid energy density, $q/\varepsilon$, to determine the parameter space where the theory remains causal. This exploration is crucial for understanding the applicability of the MIS framework to realistic physical scenarios involving significant heat fluxes.
Throughout this work, the present study focuses only on the Eckart frame as a theoretically controlled setting for isolating the causal structure associated with heat conduction in relativistic fluids (mostly relevant for finite baryon density).

For the equation of state, the present study adopts both a constant-speed-of-sound EoS, $p = c_s^2 \varepsilon$, and a realistic lattice QCD EoS, which provides a more accurate representation of strongly interacting matter at high temperatures. By comparing these two cases, the aim is to elucidate the EoS dependence of the causality constraints. To keep things simple, the present study disregards the net baryon density dependence of the speed of sound.

The present study mostly builds upon the formalism presented in \cite{Hiscock1988}, where the Eckart frame was used to analyze the stability and hyperbolicity of relativistic heat-conducting fluids. The results of this investigation provide insights into the conditions under which the MIS theory remains hyperbolic (an essential condition for well-posedness) and applicable to systems far from equilibrium, particularly for the early phase of hot and dense nuclear matter produced in high-energy heavy-ion collisions.

The remainder of this paper is organized as follows. In Sec.~\ref{sec:formalism}, the theoretical framework and governing equations are outlined. Sec.~\ref{sec:result} presents the results of the present analysis of hyperbolicity and causality for various equations of state and relaxation times. The implications of the present findings for different equations of state, including a recent lattice QCD equation of state, are also discussed. Furthermore, a brief discussion of the second-order coefficient $\beta_1$, which is an important parameter controlling the allowed values of $q/\varepsilon$ for causal wave propagation, is included. Additionally, estimates of heat flow magnitudes using the Navier-Stokes approximation are provided to assess the physical relevance of causality constraints in heavy-ion collision scenarios. Finally, the conclusions are summarized in Sec.~\ref{sec:conclusion}.

% The remainder of this paper is organized as follows. In Sec.~\ref{sec:formalism}, we outline the theoretical framework and governing equations. Sec.~\ref{sec:result} presents the results of our analysis of hyperbolicity and causality for various equations of state and relaxation times. We also discuss the implications of our findings for different equations of state, including a recent lattice QCD equation of state. Furthermore, we include a brief discussion of the second-order coefficient $\beta_1$, which is an important parameter controlling the allowed values of $q/\varepsilon$ for causal wave propagation. Finally, we summarize our conclusions in Sec.~\ref{sec:conclusion}.

Natural units with $\hbar = c = k_B = 1$ are used throughout the paper. The metric tensor is $g_{\mu\nu}=\text{diag}(1,-1,-1,-1)$.

%%%%%%%%%%%%%%%%%%%%%%%%%%%%%%%%%%%%%%%%%%%%%%%%%%%%%%%%%%%%%
\section{Formulation}
\label{sec:formalism}
For simplicity a one-dimensional fluid flow with planar symmetry is considered; the four-velocity of the fluid given by
\begin{equation}
    u^{\mu} = (\cosh\rho, \sinh\rho, 0, 0).
    \label{eq:fourvel}
\end{equation}
 Next, heat-flow $q^{\mu}$ is defined as the most general form of a space-like four-vector orthogonal to $u^\mu$ in the $x$-direction as:
\begin{equation}
q^\mu = q(\sinh\rho, \cosh\rho, 0, 0),
\end{equation}
where $q$ is a \textit{scalar coefficient} representing the magnitude and direction of heat flow (having dimension of energy density) in the spatial direction. One can verify that $q^\mu u_\mu = q(\sinh\rho \cosh\rho - \cosh\rho \sinh\rho) = 0$, as required. This demonstrates that $q^{\mu}$ is indeed space-like and orthogonal
to the fluid velocity, despite the superficial similarity in form between
the components of $u^{\mu}$ and the spatial basis vector $(\sinh\rho, \cosh\rho, 0, 0)$. In the local rest frame of the fluid ($\rho = 0$), the heat flow takes the simple form $q^\mu = (0, q, 0, 0)$, making its physical interpretation transparent. The ratio $q/\varepsilon$ appearing in the characteristic analysis (in later sections) is therefore the dimensionless measure of heat flow relative to the local energy density.

%and the four-vector heat flow \( q^{\mu} = q(\sinh\rho, \cosh\rho, 0, 0) \),
Here the fluid rapidity \( \rho \) is a function of space \((x)\) and time \((t)\). Corresponding fluid acceleration and the expansion scalars are $
a^{\mu}=u^{\nu} \partial_{\nu} u^{\mu} = \left(\cosh \rho \sinh \rho \frac{\partial \rho}{\partial t} + \sinh^2 \rho \frac{\partial \rho}{\partial x}, \cosh^2 \rho \frac{\partial \rho}{\partial t} + \cosh \rho \sinh \rho \frac{\partial \rho}{\partial x}, 0, 0 \right)
$ and  $\theta = \partial_{\nu} u^{\nu} = \sinh \rho \frac{\partial \rho}{\partial t} + \cosh \rho \frac{\partial \rho}{\partial x}
$. The co-moving derivative in this case is
$$
D\equiv u^{\mu}\partial_{\mu}=\cosh \rho \frac{\partial }{\partial t} + \sinh \rho \frac{\partial }{\partial x}.
$$
The projection tensor \(\Delta^{\mu\nu} \equiv g^{\mu\nu} -u^{\mu}u^{\nu}\) is
\begin{equation}
\Delta^{\mu\nu}=\left[\begin{array}{cccc}
-\sinh ^2 \rho & -\cosh \rho \sinh \rho & 0 & 0 \\
-\cosh \rho \sinh \rho & -\cosh ^2 \rho & 0 & 0 \\
0 & 0 & -1 & 0 \\
0 & 0 & 0 & -1
\end{array}\right].
\end{equation}
The energy-momentum tensor \(T^{\mu\nu}\) and the conserved current \(N^{\mu}\) for a dissipative fluid are given by
    % Requires: \usepackage{amsmath}
\begin{eqnarray}
      T^{\mu \nu} &=& \varepsilon u^\mu u^\nu - (p + \Pi) \Delta^{\mu \nu} + \pi^{\mu \nu} + w^\mu u^\nu + w^\nu u^\mu,
  \label{eq:energy_momentum_tensor_dissipative} \\
  N^{\mu} &=& nu^{\mu} + V^{\mu}.
\end{eqnarray}
Here, \(\varepsilon\) is the energy density, \(p\) is the thermodynamic pressure, related to \(\varepsilon\) via an equation of state (EoS), and \(\Pi\) is the bulk viscous pressure, representing the isotropic deviation of the pressure from its equilibrium value. The shear stress tensor \(\pi^{\mu \nu}\) accounts for anisotropic deviations of the stress tensor and satisfies \(\pi^{\mu \nu} u_\nu = 0\) and \(\pi^{\mu \nu} \Delta_{\mu \nu} = 0\). The energy flux four-vector \(w^\mu\) represents the flow of energy relative to the fluid's rest frame and satisfies \(w^\mu u_\mu = 0\). In the conserved current \(N^{\mu}\), \(n\) is the density of the conserved quantity (e.g., net baryon), while \(V^{\mu}\) is the particle diffusion current, representing deviations of the particle flux from the equilibrium rest frame, and it satisfies \(V^\mu u_\mu = 0\).
The definitions of these quantities in terms of energy-momentum tensor are as follows:

\begin{align*}
\varepsilon &= u_\mu u_\nu T^{\mu\nu}, \\
\Pi + p &= -\frac{1}{3} \Delta{\mu\nu} T^{\mu\nu} , \\
\pi^{\mu \nu} &= \left[\frac{1}{2} \left( \Delta^{\mu\sigma} \Delta^{\nu\tau} + \Delta^{\nu\sigma} \Delta^{\mu\tau} \right) - \frac{1}{3} \Delta^{\mu\nu} \Delta^{\sigma\tau} \right] T_{\tau\sigma}, \\
V^{\mu} &= \Delta^{\mu\nu} N_{\nu}, \\
w^\mu &= \frac{\varepsilon + p}{n} V^{\mu} + q^{\mu}, \\
n &= u_\mu N^{\mu}.
\end{align*}

The Eckart frame is adopted (where particle three current \(V^{\mu}\) vanishes) for defining the hydrodynamic four-velocity; by definition the particle diffusion \(V^{\mu}\) is zero in this frame, and the energy flow \(w^{\mu}\) coincides with the heat flow \(q^{\mu}\). It is important to note that the coefficient of thermal conductivity \(\kappa\) (defined below) may diverge for a QCD medium containing equal number of quark and antiquarks.
In the limit where the quark chemical potential \(\mu \ll T\), this divergence scales as \(\sim 1/\mu^2\). Despite the divergent \(\kappa\), the correction to the baryon current remains finite. This makes the Landau-Lifshitz frame a more suitable choice than the Eckart frame for systems produced at vanishingly small baryon chemical potential, as discussed in Ref.~\cite{PhysRevD.31.53}. For simplicity, shear and bulk viscosity are neglected in this study. The conservation equations and the Muller-Israel-Stewart (MIS) evolution equation for heat flow reduce to the following form:
\begin{eqnarray}
    Dn + n\theta &=& 0  , \label{eq:ncons}\\
    D\varepsilon +(\varepsilon + p)\theta -2 q^{\mu}a_{\mu}+\nabla_{\mu}q^{\mu} &=& 0, \\
    (\varepsilon + p) Du^{\alpha} -\nabla^{\alpha}p + q^{\mu}\partial_{\mu}u^{\alpha} + \Delta^{\alpha\nu}Dq_{\nu} + q^{\alpha}\theta &=& 0,\\
    \nonumber
    \tau_q \Delta^{\mu\nu}\dot{q}_{\nu} + q^{\mu} +\kappa \left(\nabla^{\mu}T - T \dot{u}^{\mu}\right) & & \\
    +  \left[\frac{1}{2}\kappa T^2 \partial_{\nu}\left(\frac{\tau_q}{\kappa T^2}u^{\nu}\right)q^{\mu} \right] &=& 0.
    \label{eq:heatflow}
\end{eqnarray}
Here, the relaxation time for heat flow is given by \(\tau_q = \kappa T \beta_1\), where \(\kappa\) is the coefficient of thermal conductivity, and \(\beta_1\) is a second-order transport coefficient. Following Ref.~\cite{Hiscock1988}, \(\beta_1\) is taken to be a multiple \(\lambda\) of its expression in the Israel-Stewart theory for an ultra-relativistic gas, i.e., \(\beta_1 = \lambda \frac{5}{4p}\). The parameter \(\lambda = 0\) corresponds to the acausal first-order (Navier-Stokes) theory.

For the one dimensional flow Eq.\eqref{eq:fourvel}, and EoS \(p=c_s^2\varepsilon=nT\) the conservation equations Eq.\eqref{eq:ncons} - Eq.\eqref{eq:heatflow} take the following form,
\begin{widetext}
\begin{equation}
    \cosh \rho \frac{\partial n}{\partial t} + \sinh \rho \frac{\partial n}{\partial x} + n \sinh \rho \frac{\partial \rho}{\partial t} + n \cosh \rho \frac{\partial \rho}{\partial x} = 0,
    \label{eq:ncons_newv}
\end{equation}

\begin{eqnarray}
\nonumber
   \cosh \rho \, \frac{\partial \varepsilon}{\partial t} + \sinh \rho \, \frac{\partial \varepsilon}{\partial x} + [\varepsilon(1+c_s^2) \sinh \rho + 2q \cosh \rho] \frac{\partial \rho}{\partial t} & & \\
+ \left[\varepsilon(1+c_s^2) \cosh \rho + 2q \sinh \rho\right] \frac{\partial \rho}{\partial x} + \sinh \rho \frac{\partial q}{\partial t} + \cosh \rho \frac{\partial q}{\partial x} & = & 0,
\label{eq:encons_newv}
\end{eqnarray}

\begin{eqnarray}
\nonumber
c_s^2 \sinh \rho \, \frac{\partial \varepsilon}{\partial t} + c_s^2 \cosh \rho \, \frac{\partial \varepsilon}{\partial x} + \cosh \rho \, \frac{\partial q}{\partial t} + \sinh \rho \, \frac{\partial q}{\partial x} & &\\
+ \left[\varepsilon(1+c_s^2)  \cosh \rho + 2 q \sinh \rho\right] \frac{\partial \rho}{\partial t} + \left[ \varepsilon(1+c_s^2)  \sinh \rho + 2 q \cosh \rho\right] \frac{\partial \rho}{\partial x} & = & 0,
\label{eq:momrcons_newv}
\end{eqnarray}

\begin{eqnarray}
\nonumber
\left[\frac{\sinh \rho}{\varepsilon}-\frac{10 \lambda q}{8 \varepsilon^2 c_s^2} \cosh \rho \right]\frac{\partial \varepsilon}{\partial t} + \left[\frac{\cosh \rho}{\varepsilon}- \frac{10 \lambda q}{8 \varepsilon^2 c_s^2} \sinh \rho\right]\frac{\partial \varepsilon}{\partial x} & &\\
\nonumber
 -\frac{1}{n}\left[\sinh \rho-\frac{5 \lambda q}{8\varepsilon c_s^2} \cosh \rho\right] \frac{\partial n}{\partial t} - \frac{1}{n}\left[\cosh \rho-\frac{5 \lambda q}{8\varepsilon c_s^2} \sinh \rho\right] \frac{\partial n}{\partial x} +\frac{5 \lambda }{4 \varepsilon c_s^2} \cosh \rho \frac{\partial q}{\partial t} & & \\
  + \frac{5 \lambda }{4 \varepsilon c_s^2} \sinh \rho \frac{\partial q}{\partial x} +\left[\cosh \rho + \frac{5 \lambda q}{8 \varepsilon c_s^2} \sinh \rho \right]\frac{\partial \rho }{\partial t} + \left[\sinh \rho + \frac{5 \lambda q}{8 \varepsilon c_s^2} \cosh \rho \right]\frac{\partial \rho }{\partial x} + \frac{3qn}{k\varepsilon} &=& 0.
  \label{eq:heatflow_newv}
\end{eqnarray}
\end{widetext}
Here \(T\), and \(p\) have been eliminated in terms of \(\varepsilon\) and \( n\). Equations \eqref{eq:ncons_newv} - \eqref{eq:heatflow_newv} are a set of closed equations
of the four variables \( Y^{\nu} = (n, \varepsilon, \rho, q) \). These set of equations are known to be hyperbolic (causal speed of wave propagation) if the eigenvalues are real and distinct. To calculate the characteristic speed (eigenvalues) it is usual practice to write the
above set of equations in the following form
\begin{equation}
A^{\mu \alpha}_{\nu} \partial_{\alpha} Y^{\nu} + B^{\mu} = 0,
\end{equation}
where $\nu\rightarrow $ fluid variables \((n, \varepsilon, \rho, q) \), $\mu \rightarrow$ equations \eqref{eq:ncons_newv} - \eqref{eq:heatflow_newv}, $\alpha\rightarrow$ corresponds to (\(t,x,y,z)\), since planar symmetry is assumed there is no \(y,z\) dependence.
The characteristic velocities are obtained from the determinant of the following characteristic matrix
\begin{equation}
  M=v A^{\mu t}_{\nu}-A^{\mu x}_{\nu}.
\end{equation}
The coefficients \( A^{\mu \alpha}_{\nu} \) and \( B^{\mu} \) are identified by rewriting each equation.
The explicit forms of the matrix elements \( A^{\mu \alpha}_{\nu} \) and \( B^{\mu} \) are described in detail in appendix \ref{app:matrixelem}.
Without loss of generality, the matrix \(M\) can be constructed in the fluids rest frame \(\rho=0\), in this case the analysis becomes simpler
and one has
\begin{equation}
M=\left(\begin{array}{cccc}v & 0 & -n & 0 \\ 0 & v & 2 q v-(1+c_s^2) \varepsilon & -1 \\ 0 & -c_s^2 & (1+c_s^2)\varepsilon v-2 q & v \\ \frac{5 \lambda q v}{8 \varepsilon c_s^2 n}+\frac{1}{n} & -\frac{10 \lambda q v}{8 \varepsilon^2 c_s^2}-\frac{1}{\varepsilon} & v-\frac{5 \lambda q}{8 \varepsilon c_s^2} & \frac{5 \lambda v}{4 \varepsilon c_s^2}\end{array}\right).
\end{equation}
Setting \(det(M)=0\) gives the equation for characteristic velocities
\begin{widetext}
\begin{equation}
    \left[\frac{5 \lambda}{4}\left(1+\frac{1}{c_s^2}\right)  - \frac{5 \lambda q^2}{2 c_s^2 \varepsilon^2} -1 \right] v^4 + \left[ \frac{5 \lambda q}{2 \varepsilon}\left(1-\frac{1}{c_s^2}\right) -\frac{2q}{\varepsilon}\right] v^3
+\left[c_s^2 -1 -\frac{5\lambda}{4}(1+c_s^2) + \frac{5\lambda q^2}{2c_s^2\varepsilon^2}\right] v^2 + 2\frac{q}{\varepsilon}v+ c_s^2=0.
\label{eq:fluid_charactersitic}
\end{equation}
\end{widetext}
The characteristic velocities are obtained by solving Eq.\eqref{eq:fluid_charactersitic} numerically; for a given value of \(\lambda, c_s^2\) whether four real roots exist is studied.

%%%%%%%%%%%%%%%%%%%%%%%%%%%%%%%%%%%%%%%%%%%%
\subsection{Choice of hydrodynamic frame and scope of applicability}
\label{sec:frame_choice}

Before concluding this section, it is useful to clarify the choice of hydrodynamic frame adopted in this work and the scope of applicability of the present analysis.

As discussed earlier in the Landau frame, the four-velocity is defined by the condition that the energy flow vanishes in the local rest frame, while in the Eckart frame the four-velocity is tied to the conserved charge (particle or baryon) current. For systems created in ultra-relativistic heavy-ion collisions at top RHIC and LHC energies, where the net baryon density at midrapidity is small, the Landau frame is the natural and widely adopted choice, and heat conduction does not appear as an independent dissipative current.

The situation is qualitatively different at lower beam energies, where significant baryon stopping leads to sizable net baryon densities and non-negligible baryon chemical potential. In this regime, which is of direct relevance to the ongoing RHIC Beam Energy Scan program and the upcoming FAIR and NICA experiments, transport of conserved charges and thermal conduction become essential dynamical ingredients. The Eckart frame, in which the baryon diffusion current vanishes by definition and the dissipative energy flux coincides with the heat flow, provides a physically transparent description of such systems.

The primary aim of the present work is to investigate the causal structure associated specifically with heat conduction in relativistic fluids. While it is well known that in the Landau frame the heat flow can be absorbed into baryon diffusion currents proportional to gradients of $\mu/T$, the underlying physics of thermal transport remains unchanged. Working in the Eckart frame allows the present work to analyze the causal constraints imposed by large heat fluxes directly, without simultaneously introducing additional diffusion degrees of freedom. In this sense, the Eckart-frame formulation adopted here should be viewed as a convenient and controlled theoretical setup for isolating the role of heat flow in the hyperbolicity and causality of the Muller--Israel--Stewart equations.

It is emphasized here that causality (or its violation) is a property of the constitutive structure of the hydrodynamic theory rather than of a particular frame choice. Different frames merely redistribute dissipative contributions between heat flow and diffusion currents. Therefore, any causal limitation identified in the Eckart frame is expected to have a corresponding manifestation in Landau-frame formulations at finite baryon density. From this perspective, the present analysis provides complementary insight into the limits of applicability of relativistic dissipative hydrodynamics in dense QCD matter, rather than an alternative to Landau-frame simulations commonly used at high energies.

Finally, it is worth noting that the present study is restricted to one-dimensional flow and neglects shear and bulk viscosities in order to focus sharply on the nonlinear causal structure associated with heat conduction. Extensions of this analysis to the Landau frame with explicit baryon diffusion, as well as to more general causal formulations of relativistic hydrodynamics, will be addressed in future work.

%%%%%%%%%%%%%%%%%%%%%%%%%%%%%%%%%%%%%%%%%%%%%%%%%%

\begin{figure*}
\includegraphics[scale=0.5]{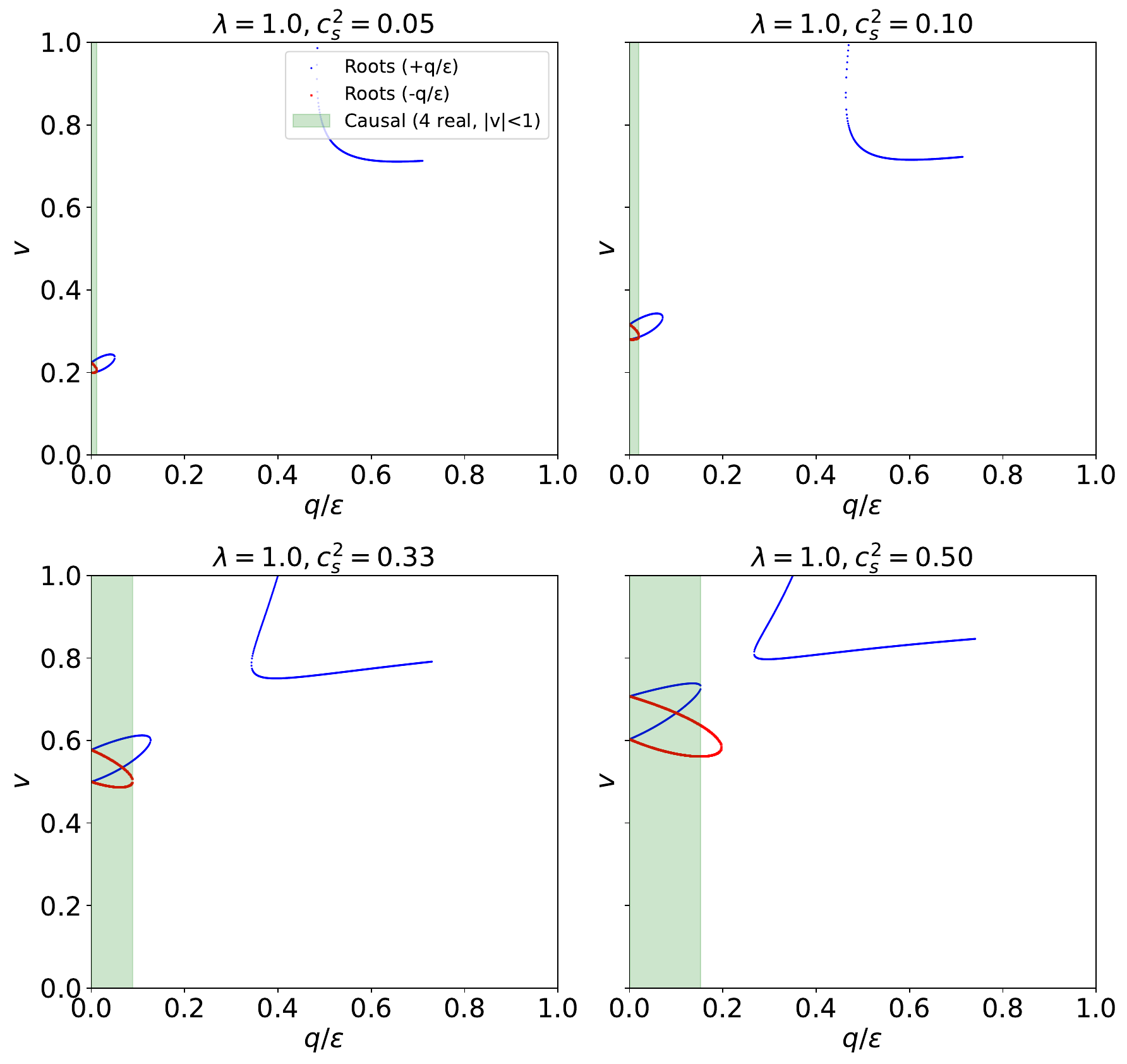}% Here is how to import EPS art
\caption{\label{fig:wide} Characteristic velocities \(v\) obtained from the numerical solution of Eq.~\eqref{eq:fluid_charactersitic} shown as a function of \(q/\varepsilon\), for \(\beta_1= \frac{5}{4p}\) and four different $c_s^2$=0.05, 0.10, 0.33, and 0.5. The shaded green region indicates the causal hyperbolic region.}
\label{fig:chracter_lambda_1.0}
\end{figure*}

%%%%%%%%%%%%%%%%%%%%%%%%%%%%%%%%%%%%%%%%%%%%%%%
%% RESULT                                %%%%%%
%%%%%%%%%%%%%%%%%%%%%%%%%%%%%%%%%%%%%%%%%%%%%%%
\section{Results}
\label{sec:result}

The causal structure of the MIS theory is analyzed by numerically solving the characteristic equation, Eq.~\eqref{eq:fluid_charactersitic}, for the propagation velocities $v$. The system is hyperbolic and causal when all four roots are real and subluminal ($|v| \le 1$). The causal parameter space is investigated as a function of the relaxation time parameter, $\lambda$ (where $\beta_1 = \lambda \frac{5}{4p}$), the equation of state (EoS), and the normalized heat flux, $q/\varepsilon$.

For the EoS, several constant values for the squared speed of sound, $c_s^2$, are considered. While a relativistic massless gas corresponds to $c_s^2=1/3$, stiffer cases up to $c_s^2=0.5$ are also analyzed, as such high values may be reached in extremely dense systems~\cite{KapustaGale2006, Weinberg1972, Bedaque2015}.

\begin{figure*}[t]
\centering
\includegraphics[scale=0.5]{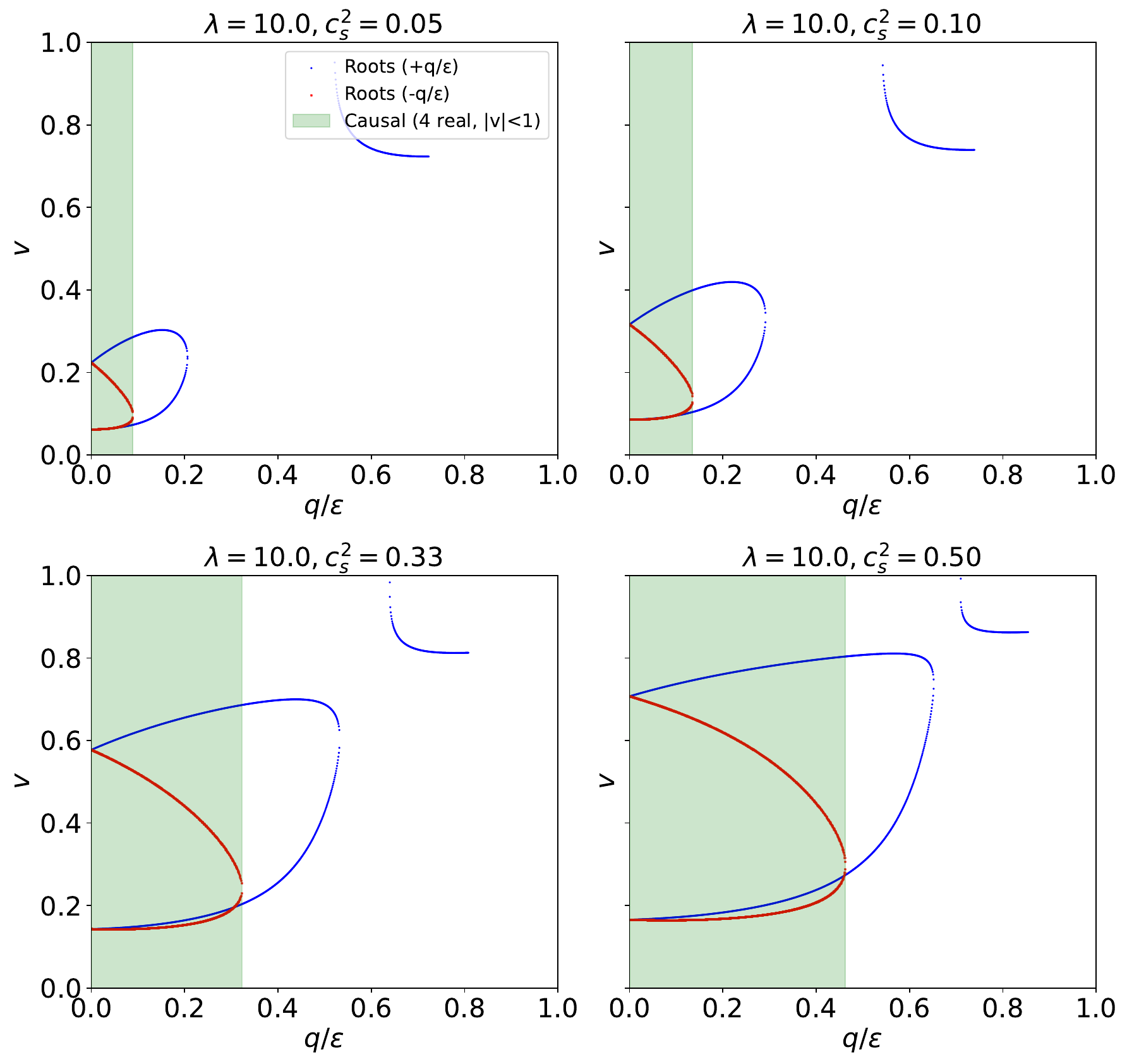}
\caption{Characteristic velocities $v$ as functions of $\frac{q}{\varepsilon}$ for $\lambda = 10$ and four different values of $c_s^2$: $0.05, 0.1, 0.33, 0.5$. The causal regions, marked by the shaded green area, show a significant enhancement in the range of $\frac{q}{\varepsilon}$ compared to smaller $\lambda$ cases.}
\label{fig:chracter_lambda_10.0}
\end{figure*}

Figures~\ref{fig:chracter_lambda_1.0} and \ref{fig:chracter_lambda_10.0} show the characteristic velocities for $\lambda=1$ and $\lambda=10$. Two key trends are immediately apparent. First, for a fixed $\lambda$, a stiffer EoS (larger $c_s^2$) consistently expands the range of $|q/\varepsilon|$ for which causality is maintained. Second, comparing the two figures shows that a larger relaxation time (increasing $\lambda$ from 1 to 10) significantly enlarges the causal domain. For instance, in the ultra-relativistic limit ($c_s^2=1/3$), the causal region is limited to $|q/\varepsilon| \lesssim 0.18$ for $\lambda=1$ (consistent with Hiscock and Lindblom~\cite{Hiscock1988}), but expands to $|q/\varepsilon| \lesssim 0.25$ for $\lambda=10$.

\begin{figure*}[t]
\centering
\includegraphics[width=0.3\textwidth]{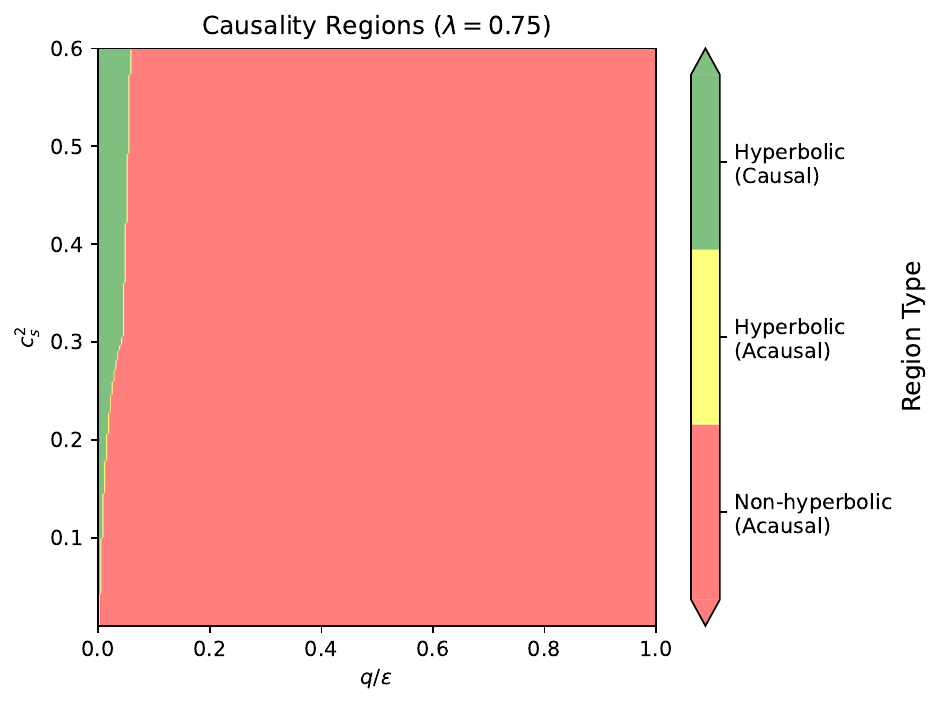}
\includegraphics[width=0.3\textwidth]{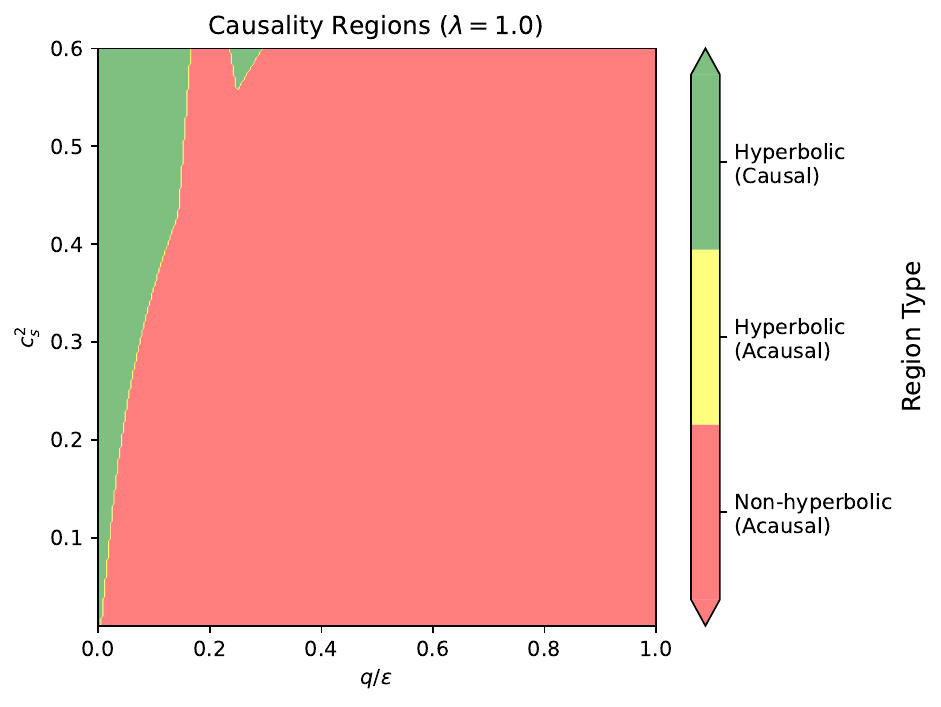}
\includegraphics[width=0.3\textwidth]{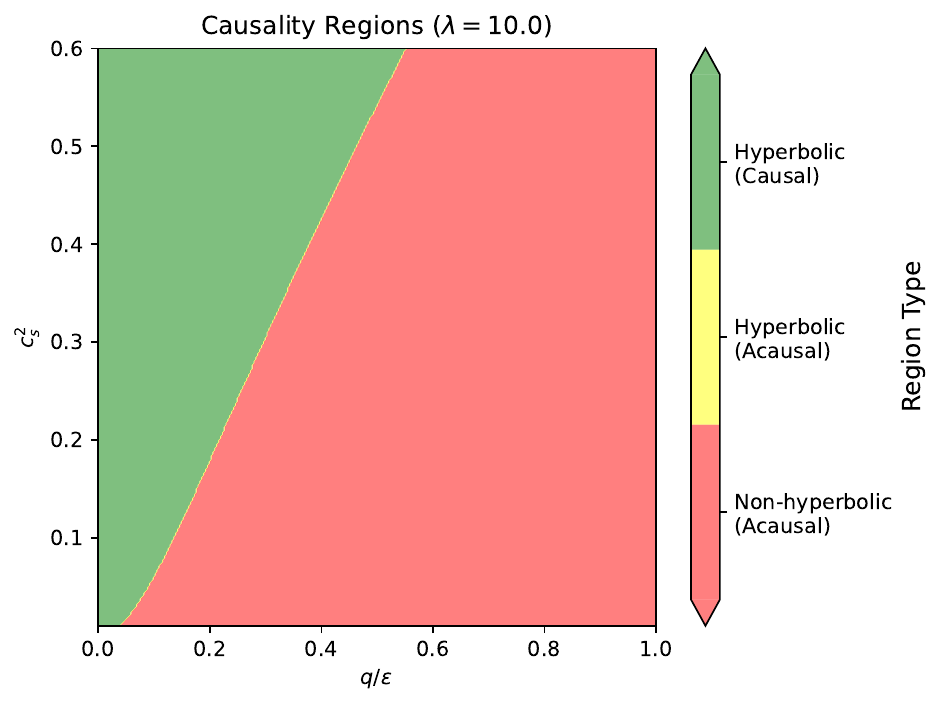}
\caption{Causality regions in the ($q/\varepsilon$, $c_s^2$) plane for different values of the relaxation parameter, $\lambda = 0.75, 1.0,$ and $10.0$. The green (Hyperbolic/ Causal), yellow (Hyperbolic/Acausal), and red (Non-hyperbolic/Acausal) regions denote the parameter space where the characteristic velocities are all real and subluminal, all real but at least one is superluminal, and not all real, respectively.}
\label{fig:2d_causality}
\end{figure*}

To visualize these dependencies more globally, Fig.~\ref{fig:2d_causality} maps the causality regions in the ($q/\varepsilon$, $c_s^2$) plane for $\lambda = 0.75, 1.0,$ and $10.0$. These plots confirm that the causal domain (green shaded area) is highly sensitive to both the EoS and the relaxation time. The region of  causal evolution shrinks dramatically for softer equations of state and smaller values of $\lambda$. This strong interplay shows the important role of the second-order transport coefficients and the EoS in maintaining a well-posed hydrodynamic theory, particularly in regimes with large temperature gradients where the heat flux may be significant.

\subsection{\(c_s^2(\varepsilon)\) from Lattice QCD result}
Up until now, the results for a temperature-independent constant speed of sound have been presented. But in reality, the speed of sound depends on both temperature and density. Here, only the temperature dependence of \(c_s^2\) is considered; the inclusion of density dependence could be done in a similar fashion.

\begin{figure}
    \centering
    \includegraphics[width=1.0\linewidth]{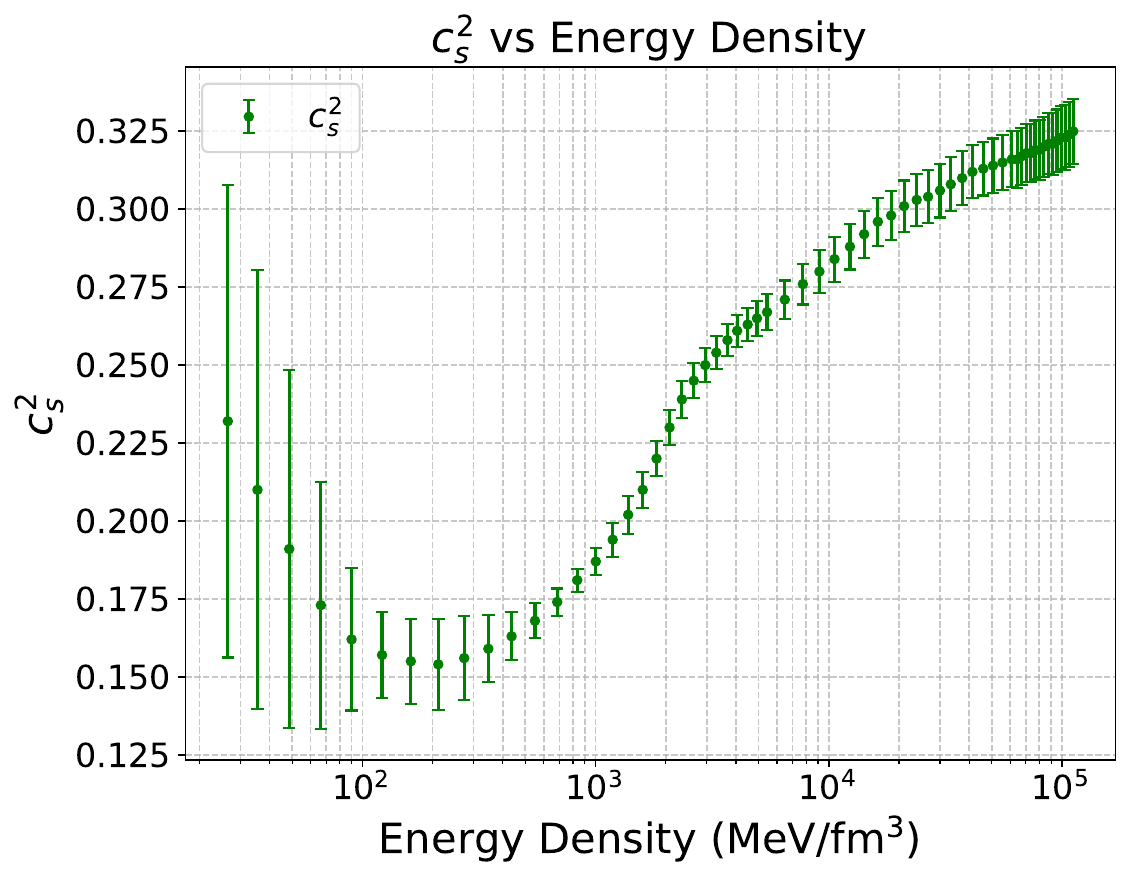}
    \caption{Lattice QCD results for squared speed of sound \(c_s^2\) as a function of energy density. }
    \label{fig:c_s2LQCD}
\end{figure}

To study the realistic scenario of a temperature-dependent \(c_s^2\), LQCD results from Ref.~\cite{Borsanyi2014} are used. The tabulated data for thermodynamic variables provided in \cite{Borsanyi2014} for the \(N_f = 2+1\) QCD equation of state as a function of temperature are used and converted to express \(c_s^2\) as a function of \(\varepsilon\) for use in the calculation. The corresponding plot for \(c_s^2\) as a function of \(\varepsilon\) is shown in Fig.~\eqref{fig:c_s2LQCD}. As can be seen from Fig.~\eqref{fig:c_s2LQCD}, \(c_s^2\) exhibits a non-monotonic variation, with a dip around \(\approx 130\) MeV and approaching the massless limit \(c_s^2 = \frac{1}{3}\) from below in the high-temperature regime. This non-trivial temperature dependence of the speed of sound has interesting implications for the corresponding causal region.

\begin{figure}
    \centering
    \includegraphics[width=1.0\linewidth]{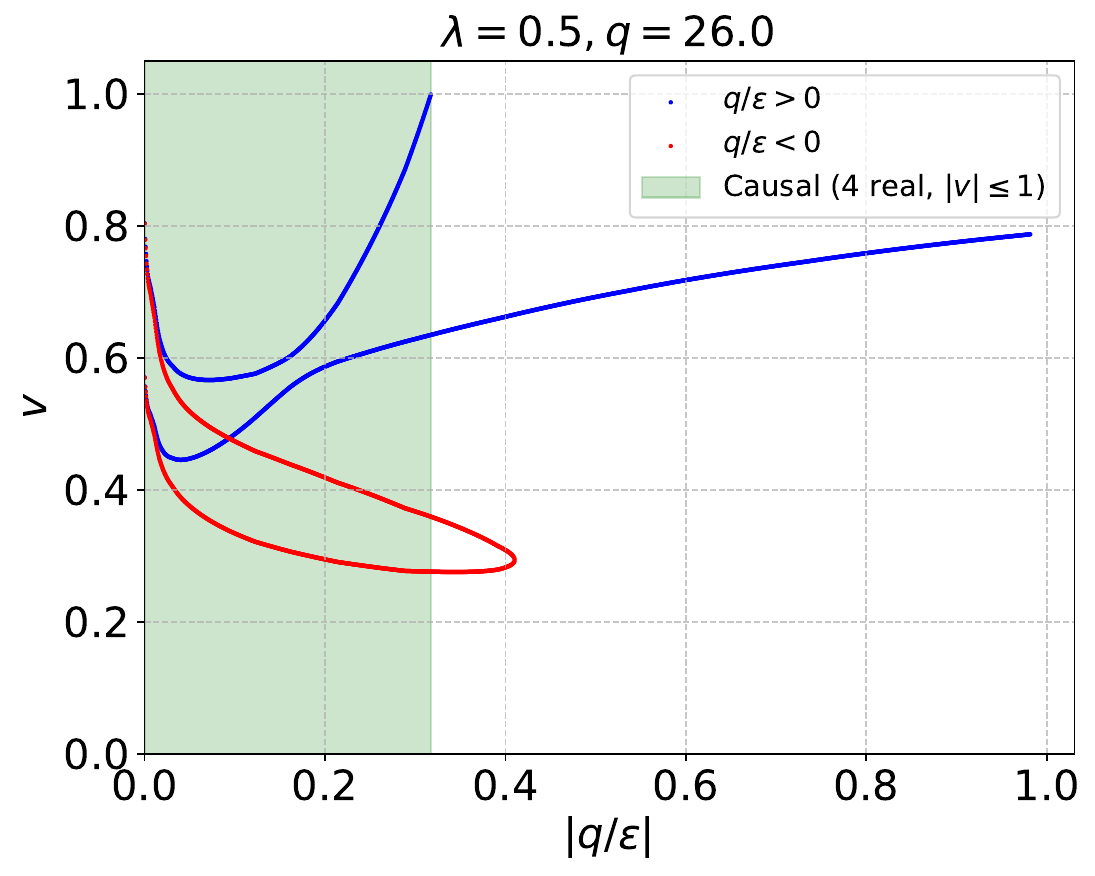}
    \includegraphics[width=1.0\linewidth]{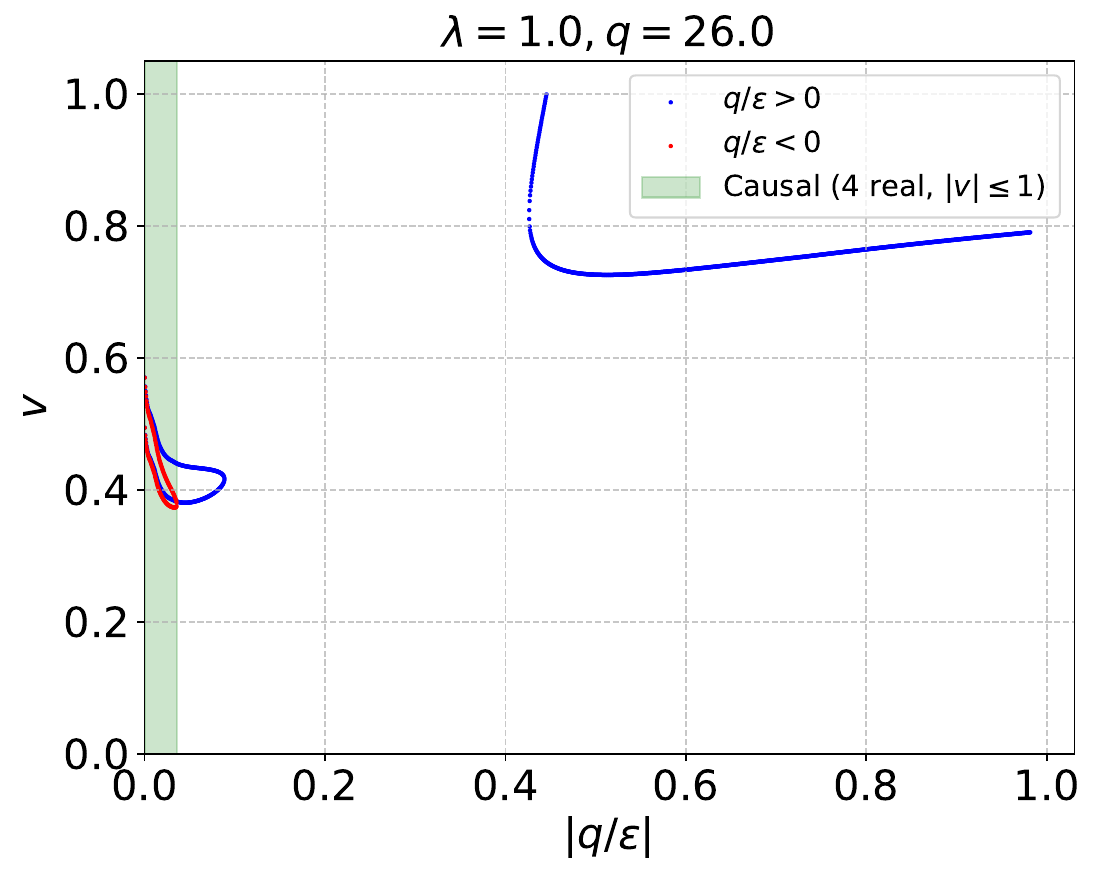}
    \includegraphics[width=1.0\linewidth]{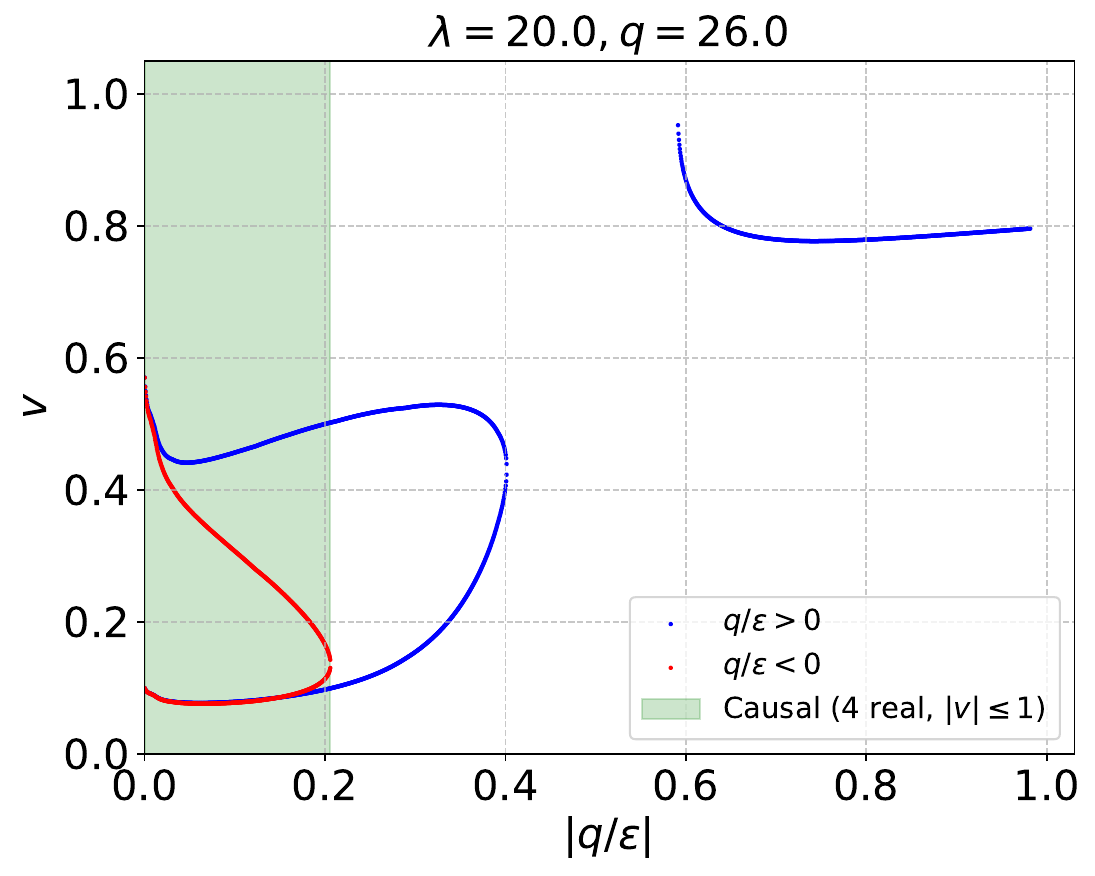}
    \caption{Characteristic velocities $v$ as functions of $q/\varepsilon$ for three different values of $\lambda = 0.5, 1.0,$ and $20.0$, calculated using a lattice QCD equation of state. The plots show the results for positive $q/\varepsilon$, with the causal hyperbolic region highlighted by green rectangles.}
    \label{fig:LQCDresult}
\end{figure}

 The characteristic velocities \(v\) for three different values of the relaxation parameter \(\lambda = 0.5, 1.0,\) and \(20.0\) are shown in Fig.~\ref{fig:LQCDresult}  for the temperature-dependent speed of sound as shown in Fig.~\ref{fig:c_s2LQCD}.

From the top panel of Fig.~\ref{fig:LQCDresult} corresponding to \(\lambda = 0.5\), one observes that the characteristic velocities exhibit significant variation over a relatively wide range of \(q/\varepsilon\), with regions of multiple real roots indicating causal propagation. However, the structure is considerably more intricate compared to the constant \(c_s^2\) case, with more pronounced non-linearities in the velocity profile. The causal region appears to extend into both positive and negative values of \(q/\varepsilon\), though the subluminal condition \(v < 1\) may be violated in some regions, indicating instability for extreme gradients.

The middle panel shows the result for \(\lambda = 1.0\). Here, the causal region is more tightly constrained, and the structure of the velocity curves is comparatively simpler than for \(\lambda = 0.5\). One finds that for small values of \(q/\varepsilon\), all four roots remain real and subluminal, while beyond a certain threshold the number of real roots reduces, signaling a loss of causality. Compared to the constant \(c_s^2\) scenario, the causal region is somewhat suppressed, likely due to the dip in \(c_s^2\) around the crossover region in the LQCD equation of state.

The bottom panel corresponds to a much larger relaxation parameter, \(\lambda = 20.0\), representing a system with significantly delayed relaxation to equilibrium. In this case, a complex structure of the characteristic velocities is again observed, but the causal region appears to broaden compared to the \(\lambda = 1.0\) case, consistent with the trend that increasing \(\lambda\) tends to support causality over a wider range of heat flux. However, similar to the \(\lambda = 0.5\) case, the high degree of non-linearity in the velocity structure can introduce regions where some roots may exceed the light cone or become complex.

Overall, the figure highlights that the use of a realistic, temperature-dependent speed of sound introduces substantial modifications to the causal structure compared to the idealized constant \(c_s^2\) case. In particular, the dip in \(c_s^2\) near \(T \approx 130\) MeV seen in LQCD results leads to a narrowing of the causal window for intermediate energy densities.

\subsubsection{Discussion on $\beta_1$}
As is clear from the above discussion, the causal parameter space is quite sensitive to the second-order transport coefficient \(\beta_1\) appearing in MIS theory for heat flow. It is worthwhile to discuss its physical values for the temperature ranges of interest. Kinetic theory calculation based on Grad's fourteen moment method for a Boltzmann gas gives the following values for
\(\beta_1 P\) for ultra-relativisitic ($z\ll 1$), non-relativistic ($z\gg 1$) and in the intermediate ($z\sim 1$) regions

$\beta_1 P =
\begin{cases}
\dfrac{5}{4}, ~~~~ \text{for ~} z \ll 1 ~ \text{(Ultra-relativistic)} \\
\\
\dfrac{2}{5} z, ~~~~\text{for } z \gg 1 \quad \text{(Non-relativistic)}\\
\left( \dfrac{\gamma - 1}{\gamma} \right)^2 \cdot \dfrac{z}{y} \cdot \left(5 y^2 - X \right), ~~ \text{otherwise} \\
\end{cases}$
where
\begin{eqnarray}
 y &=& \frac{K_3(z)}{K_2(z)}, \\
 X &=& z^2 \left(1 + \frac{5 y}{z} - y^2 \right), \\
 \gamma &=& \frac{X}{X - 1},
\end{eqnarray}
 $z=\frac{m}{T}$, and $K_n(z)$ denotes the modified Bessel function of the second kind of order $n$.

\begin{figure}
    \centering
    \includegraphics[width=1.0\linewidth]{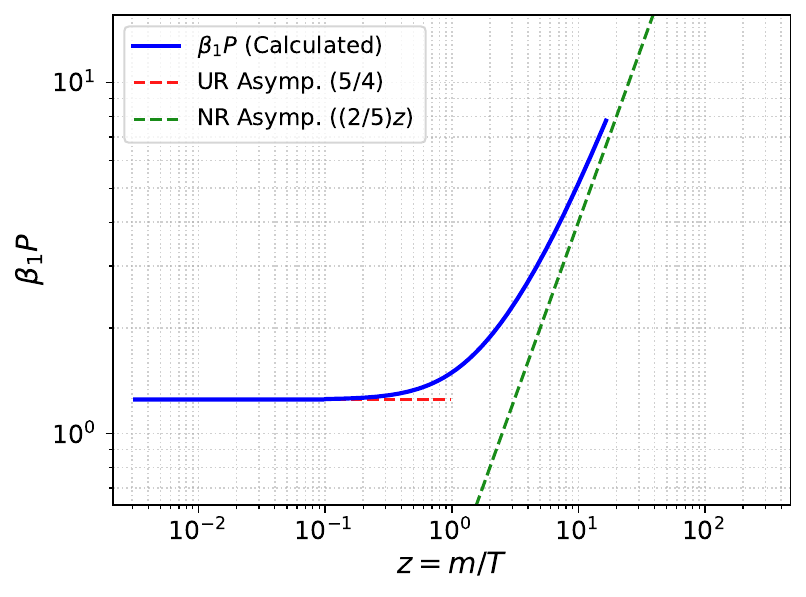}
    \caption{Second order coefficient $\beta_1$ times pressure as a function of $z=\frac{m}{T}$. The red and green dashed lines are ultra-relativistic and non-relativistic limits respectively, and the solid blue line is the exact calculation.}
    \label{fig:Beta1p}
\end{figure}

Fig.\eqref{fig:Beta1p} shows the variation of $\beta_1 P$ as a function of $z$.
As can be seen lower temperature ( larger $z$) corresponds to larger $\beta_1$
which implies larger relaxation time for heat flow. On the other hand the large temperature limit gives \(\beta_1\propto \frac{1}{p}\) that implies \(\beta_1\) becomes smaller in the high-temperature limit.

\subsubsection{Estimation of heat flow in the Navier-Stokes limit}
At this point, it is interesting to obtain an order-of-magnitude estimate of the heat flow values that occur in RHIC and lower-energy collisions to assess the causal zones allowed by the MIS theory. Consider the simplest Navier-Stokes approximation of the heat flow equation ($\tau_q \rightarrow 0$) that implies the heat flux responds almost instantaneously to changes in the thermodynamic gradients, thereby eliminating the ``memory'' effects that are characteristic of the Israel-Stewart formulation.
Eq.~\eqref{eq:heatflow} in this limit takes the following form:
\begin{equation}
      q^{\mu} +\kappa \left(\nabla^{\mu}T - T \dot{u}^{\mu}\right) = 0.
\label{eq:NSheat}
\end{equation}
The above equation states that the heat flux is directly proportional to the spatial gradient of temperature and an additional relativistic term involving the fluid's acceleration. For further simplification, the zeroth-order equation of motion is used and $\dot{u}^{\mu}=\frac{\nabla^{\mu} p}{(\varepsilon +p)}$ is set in Eq.~\eqref{eq:NSheat}.
This gives the Navier-Stokes expression for heat flow as
\begin{equation}
  q^{\mu} = - \kappa \left(\nabla^{\mu}T - \frac{T}{(\varepsilon +p)}\nabla^{\mu} p\right).
\label{eq:NSheat2}
\end{equation}
If the fluid acceleration is disregarded by setting $\dot{u}^{\mu}=0$ in Eq.~\eqref{eq:NSheat}, the simplest approximation $q^{\mu} = - \kappa \nabla^{\mu}T$ is obtained. Below, the individual and combined contributions to the total heat flux from the temperature gradient and pressure gradient will be shown.
The ratio of the magnitude of the spatial part of the heat flow vector to energy density, i.e., $|\mathbf{q}|/\varepsilon$, is of primary interest. To obtain $|\mathbf{q}|$, appropriate values for $\kappa$ must be provided.
To the best of current knowledge, first-principle lattice QCD results for thermal conductivity $\kappa$ are not currently available, mainly due to theoretical constraints and technical challenges involved in extracting the necessary correlation functions, especially since thermal conductivity is only an independent transport coefficient in the presence of conserved charges. However, several QCD-inspired model calculations (notably BGK/quasiparticle models) do provide estimates and qualitative insights into the behavior of thermal conductivity in hot and dense QCD matter. For example, $\kappa$ has been calculated from relativistic kinetic theory and the NJL model in \cite{PhysRevD.31.53,Chakrabarty1985,Mitra:2017sjo,KadamPRD114001,Greif:2013bb,Marty:2013ita}, etc.

However, all these studies provided varying temperature dependence of $\kappa$. For example, in \cite{PhysRevD.31.53}, the expression for $\kappa$ in the small $\mu/T$ and symmetric quark-antiquark $|n_q -n_{\bar{q}}| \ll (n_q + n_{\bar{q}})$ limit is
\begin{equation}
    \kappa = \frac{16}{9} \frac{K_{\mathrm{SB}}^2}{N_c N_f}\left(\tau_q T\right) \frac{T^4}{\mu^2},
    \label{eq:kappaGyulassy}
\end{equation}
where
$K_{\mathrm{SB}}=\left[\left(N_c^2-1\right)+7 N_c N_f / 8\right] \pi^2 / 15$
is the Stefan-Boltzmann constant for $N_f$ flavors and $N_c$ colors, and $\tau_q T$ was shown to be proportional to $1/(\alpha_s^2 \ln (1/\alpha_s))$.
According to the above formulation, $\kappa/T^2$ should increase as a function of temperature for a given $\mu/T$, since the strong coupling $\alpha_s$ decreases as a function of temperature. For simplicity, the calculation is performed for a system with a characteristic radius of $R = 5$ fm and central temperature of $T_0=200$ MeV. The temperature is further assumed to vary linearly along the radial direction: $T(r) = T_0 \left(1 - \frac{r}{R} \right)$. Using Eq.~\eqref{eq:kappaGyulassy}, $\kappa$ is calculated for different $\mu/T$ values as given in Table~\ref{tab:mubyTvsKappaT200}, where $N_c=3$ and $N_f=2$ with the one-loop result for running $\alpha_s$ are used in the calculation. The heat flow for such a linear radial dependence of temperature and for $\mu/T = 0.1$ and central temperature $T_0=200$ MeV is $|{\bf{q}}_{\nabla T}| \approx 711$ GeV/fm$^3$. Here only the simplest approximation driven by the temperature gradient ($q^{i} \approx - \kappa \nabla^{i}T$) is considered. The thermal conductivity is taken to be $\kappa = 6.9\times 10^{8}$ MeV$^2$. For the thermodynamic quantities, the Wuppertal-Budapest lattice QCD equation of state~\cite{Borsanyi2014} is used.

%%%%%%%%%%%%%%%%%%%%%%%%%%%%%%%%%%%%%%%%%%%%%%%%%%%%%%
% Next, we include the correction term from Eq.~\eqref{eq:NSheat2}. This term, which accounts for the pressure gradient, provides a contribution that opposes the primary heat flow. For the given parameters, its magnitude is calculated to be $|{\bf{q}}_{\nabla p}| \approx \SI{108}{\giga\electronvolt\per\femto\meter\cubed}$.
% The inclusion of this correction term results in a total heat flux of $|q_{\text{total}}| \approx \SI{603}{\giga\electronvolt\per\femto\meter\cubed}$. This shows that the pressure gradient provides a significant, non-trivial contribution, reducing the magnitude of the heat flow by approximately \textbf{15\%} compared to the simple gradient approximation.
% The resulting ratio of the total heat flux to the central energy density is found to be $|{\bf{q}}_{\text{total}}| / \varepsilon_0 \approx 330$ where we take \(\varepsilon_0 \sim \) $\SI{1.8}{\giga\electronvolt\per\femto\meter\cubed}$ at \(T=200\) MeV temperature. Needless to say that $|{\bf{q}}_{\text{total}}| / \varepsilon_0 \approx 330$ is considered to be too large a value to be realistic.
%%%%%%%%%%%%%%%%%%%%%%%%%%%%%%%%%%%%%%%%%%%%%%%%%
Next, the correction term from Eq.~\eqref{eq:NSheat2} is included. This term, which accounts for the pressure gradient, provides a contribution that opposes the primary heat flow. For the given parameters, its magnitude is calculated to be $|{\bf{q}}_{\nabla p}| \approx 108$ GeV/fm$^3$.
The inclusion of this correction term results in a total heat flux of $|{\bf{q}}_{\text{total}}| \approx 603$ GeV/fm$^3$. This shows that the pressure gradient provides a significant, non-trivial contribution, reducing the magnitude of the heat flow by approximately \textbf{15\%} compared to the simple gradient approximation.
The resulting ratio of the total heat flux to the central energy density is found to be $|{\bf{q}}_{\text{total}}| / \varepsilon_0 \approx 330$, where $\varepsilon_0 \approx 1.8$ GeV/fm$^3$ at $T=200$ MeV temperature is taken. It is evident that $|{\bf{q}}_{\text{total}}| / \varepsilon_0 \approx 330$ is too large a value to be physically realistic.

%%%% New addition after PRC referee %%%%%%%%%%
 Such extreme values of heat flux raise fundamental concerns
beyond causality alone. As demonstrated in ~\cite{HawkingEllis1973, HiscockOlson1989} ,
when heat flux becomes sufficiently large in the Eckart frame, the stress-energy
tensor violates the weak energy condition---specifically, when
$|q|/(\varepsilon+p) > 1/2$, some observers will measure negative energy
densities. For the parameters considered here, $|q|/(\varepsilon+p) \approx 200$--$400$,
vastly exceeding this fundamental physical bound. This violation indicates
that the underlying fluid description itself becomes pathological, independent
of causality considerations.

%%%%%%%%%%%%%%%%%%%%%%%%%%%%%%%%%%%%%%%%%%%%%%%%%

\begin{table}[htb]
\centering
\begin{ruledtabular}
\begin{tabular}{cc}
$\mu/T$ & $\kappa$ (MeV$^2$) \\
\hline
0.10 & $6.92 \times 10^{8}$ \\
%0.20 & $1.73 \times 10^{8}$ \\
0.30 & $7.69 \times 10^{7}$ \\
%0.40 & $4.33 \times 10^{7}$ \\
0.50 & $2.77 \times 10^{7}$ \\
\end{tabular}
\end{ruledtabular}
\caption{Heat conductivity $\kappa$ as a function of $\mu/T$ for $N_c=3$, $N_f=2$ at a fixed temperature of $T=200$ MeV.}
\label{tab:mubyTvsKappaT200}
\end{table}
Similar calculations for a somewhat lower central temperature $T=150$ MeV and $\mu/T = 0.1$, but keeping other parameters fixed, result in a somewhat larger $|{\bf{q}}_{\text{total}}| / \varepsilon_0 \approx 811$ (where $\kappa \approx 3.31 \times 10^{8}$ MeV$^2$ is taken from Table~\ref{tab:mubyTvsKappaT150}). The apparent increase in $|{\bf{q}}_{\text{total}}| / \varepsilon_0$ for lower temperature is due to the fact that $\varepsilon \sim T^{4}$, whereas $\kappa \sim T^2$.

Larger values of $\mu/T$ will not lead to a significant decrease in the heat flow because $\kappa$ changes by only one order of magnitude as $\mu/T$ increases from 0.1 to 0.5. Hence, this will not give any acceptable values for $|{\bf{q}}|/\varepsilon$ either. On the other hand, if $\kappa$ is hypothetically taken to be two orders of magnitude smaller, i.e., $\kappa \sim 10^{5}$ MeV$^2$, that will give $|{\bf{q}}_{\text{total}}|/\varepsilon \approx 0.24$ at $T=150$ MeV, which is somewhat reasonable but still considered to be a large value.

Since it is clear that the large heat flow originates mainly due to the large value of $\kappa$, it is worthwhile to check its value from other theoretical model studies.
A more recent calculation \cite{Mitra:2017sjo} based on semi-classical transport theory using a quasi-particle description of a system of quarks, antiquarks, and gluons with a realistic QCD equation of state gives a somewhat different temperature dependence (for non-zero quark chemical potentials):
$$
\kappa =
\begin{cases}
\sim 150 \times T^2, & \text{for } T/T_c \sim 1.5, \\
\\
\sim 25 \times T^2, & \text{for } T/T_c \sim 5.  \\
\end{cases}
$$
Taking $T_c\sim 150$ MeV, this corresponds to $\kappa \sim 7.5\times 10^{6}$ MeV$^2$ at $T=225$ MeV and $\kappa \sim 1.4 \times 10^{7}$ MeV$^2$ at $T=750$ MeV. These values are comparable to what was reported in Tables~\ref{tab:mubyTvsKappaT200} and~\ref{tab:mubyTvsKappaT150}. The thermal conductivity coefficient was calculated in \cite{Greif:2013bb} using a numerical solution of the relativistic Boltzmann equation for a classical gas of massless particles with elastic binary collisions and constant isotropic cross-section, and in Ref.~\cite{Marty:2013ita} using the NJL model. These results are in good agreement with the above reported values.

%%%%%%%%%%%%%%%%%%%%%%%%%%%%%%%%%%%%%%%%%%%%%%%%%%%%%%%%%%
\begin{table}[htb]
\centering
\begin{ruledtabular}
\begin{tabular}{cc}
$\mu/T$ & $\kappa$ (MeV$^2$) \\
\hline
0.10 & $3.31 \times 10^{8}$ \\
0.30 & $3.68 \times 10^{7}$ \\
0.50 & $1.32 \times 10^{7}$\\
\end{tabular}
\end{ruledtabular}
\caption{Heat conductivity $\kappa$ as a function of $\mu/T$ for $N_c=3$, $N_f=2$ at a fixed temperature of $T=150$ MeV.}
\label{tab:mubyTvsKappaT150}
\end{table}
%%%%%%%%%%%%%%%%%%%%%%%%%%%%%%%%%%%%%%%%%%%%%%%%%%%%%%%%%%
% At this point, we would like to mention that the decisive factor for large heat-flow \(\kappa\) used here is obtained from kinetic theory calculation; a first principle calculation of \(\kappa\) from Lattice QCD  is required to get an accurate estimate of the heat flow. We should also note that the above calculation is done for Navier-Stokes limit, and as we know that according to the MIS theory, heat flow is a dynamical variable like other dissipative quantities and will follow a transient behavior where its magnitude during the initial times could be very different (large) from the  Navier stokes limit and only approaches it in the late time. That implies causality criteria in heavy-ion collisions related to heat-flow is on the brink or perhaps broken, a fact which is also shown in numerical studies \cite{Plumberg2022}. The question is more relevant for smaller systems with larger temperature gradients.

%%%%%%%%%%%%%%%%%%%%%%%%%%%%%%%%%%%%%%%%%%%%%%%%%%%%%%%%
At this point, it is worth mentioning that the decisive factor for large heat flow is the value of $\kappa$ used here, which is obtained from kinetic theory calculations; a first-principle calculation of $\kappa$ from lattice QCD is required to obtain an accurate estimate of the heat flow. It should also be noted that the above calculation is performed in the Navier-Stokes limit, and as is known from the MIS theory, heat flow is a dynamical variable like other dissipative quantities and will follow a transient behavior where its magnitude during the initial times could be very different (and large) from the Navier-Stokes limit and only approaches it at late times. This implies that causality criteria in heavy-ion collisions related to heat flow are on the brink or perhaps broken, a fact which is also shown in numerical studies \cite{Plumberg2022} but for non-zero shear and bulk viscosities.
This question is more relevant for smaller systems, such as proton-proton collisions or peripheral heavy-ion collisions, with larger temperature gradients.

% \section{Conclusion}
% \label{sec:conclusion}
% In conclusion, we have analyzed the causality conditions for a one-dimensional relativistic fluid with heat flow. By solving the characteristic equation numerically, we determined the ranges of $\frac{q}{\varepsilon}$ for which the system exhibits hyperbolic behavior, ensuring causal propagation of signals. Our results demonstrate that both the equation of state and the relaxation parameter significantly influence the causal structure of the system.
% To maintain causality, the eigenvalues of the characteristic equations must be real and respect subluminal signal propagation conditions. The constraints on relaxation times for dissipative stresses are crucial in this regard, as improper values may introduce acausal modes or instabilities. The exact constraints require solving for the eigenvalues explicitly and analyzing their dependence on the transport coefficients.
% In the Navier-Stokes limit we further estimate the ratio of heat flow to the fluid central energy density under certain simplified assumptions and the corresponding values turned out to be too large to be physically realistic for the temperature range and \(\kappa\) (obtained from other theoretical model studies) applicable to RHIC and lower energy heavy-ion collisions. We believe a first principle calculation of \(\kappa\) from LQCD is needed for better constraint causal space.

% Future work could explore the implications of these findings for higher-dimensional flows and more realistic equations of state for larger baryon chemical potentials.
\section{Conclusion}
\label{sec:conclusion}
In conclusion, the causality conditions for a one-dimensional relativistic fluid with heat flow have been analyzed. By solving the characteristic equation numerically, the ranges of $\frac{q}{\varepsilon}$ for which the system exhibits hyperbolic behavior were determined, ensuring causal propagation of signals. The present results show that both the equation of state and the relaxation time significantly influence the causal structure of the system.

To maintain causality, the eigenvalues of the characteristic equations must be real and respect subluminal signal propagation conditions. The constraints on relaxation times for dissipative stresses are crucial in this regard, as improper values may introduce acausal modes or instabilities. The exact constraints require solving for the eigenvalues explicitly and analyzing their dependence on the transport coefficients.

In the Navier-Stokes limit, the ratio of heat flow to the fluid central energy density was further estimated under certain simplified assumptions. The corresponding values turned out to be unrealistically large ($|{\bf{q}}_{\text{total}}| / \varepsilon_0 \approx 330$ for $T=200$ MeV and $\approx 811$ for $T=150$ MeV) for the temperature range and thermal conductivity $\kappa$ values obtained from theoretical model studies applicable to RHIC and lower-energy heavy-ion collisions. These results suggest that either the current estimates of thermal conductivity are overestimated, or that the applicability of dissipative fluid dynamics breaks down in these extreme conditions.
%%%%%%%%% New addition after PRC referee %%%%%%
Furthermore, it is found that the estimated heat flux values violate
not only causality bounds but also the weak energy condition
($|q|/(\varepsilon+p) \approx 200$--$400 \gg 1/2$), indicating a fundamental
breakdown of the fluid description at these extreme gradients.
%%%%%%%%%%%%%%%%%%%%%%%%%%%%%%

The present analysis reveals that the pressure gradient correction provides a non-trivial contribution, reducing the heat flow magnitude by approximately 15\% compared to the simple temperature gradient approximation. However, even with this correction, the heat flow remains orders of magnitude larger than what would be considered physically reasonable.

It is argued here that first-principle calculations of $\kappa$ from lattice QCD are essential for better constraining the causal parameter space and providing reliable predictions for heavy-ion collision phenomenology.

As discussed in detail in Sec.~\ref{sec:frame_choice}, while the present analysis is performed in the Eckart frame for physical and pedagogical reasons, the
causal constraints identified here are expected to have corresponding manifestations in Landau-frame formulations at finite baryon density.

Future work could explore the implications of these findings for higher-dimensional flows, more realistic equations of state at larger baryon chemical potentials, and the transition from the early non-equilibrium phase to the hydrodynamic regime in heavy-ion collisions. Additionally, investigating the role of other transport coefficients and their interplay with thermal conductivity would provide a more complete picture of causal parameter space in relativistic heavy-ion collisions in the non-linear regime.

\begin{acknowledgments}
V.R. acknowledges the financial support from Anusandhan National Research Foundation~(ANRF), India through Core Research Grant , CRG/2023/001309.  V.R. also acknowledges
Department of Atomic Energy~(DAE), India for financial support.
\end{acknowledgments}

\appendix

\section{Matrix Elements of \( A^{\mu \alpha}_{\nu} \)}
\label{app:matrixelem}

Listed below are the non-zero components of \( A^{\mu \alpha}_{\nu} \), organized by equation index \( \mu \), derivative index \( \alpha \), and fluid variable index \( \nu \).

\vspace{1em}
\noindent
\textbf{From Eq.~\eqref{eq:ncons_newv}:}
\begin{align*}
A^{1t}_n &= \cosh \rho, \\
A^{1x}_n &= \sinh \rho, \\
A^{1t}_\rho &= n \sinh \rho, \\
A^{1x}_\rho &= n \cosh \rho.
\end{align*}

\vspace{1em}
\noindent
\textbf{From Eq.~\eqref{eq:encons_newv}:}
\begin{align*}
A^{2t}_\varepsilon &= \cosh \rho, \\
A^{2x}_\varepsilon &= \sinh \rho, \\
A^{2t}_\rho &= \varepsilon (1 + c_s^2) \sinh \rho + 2q \cosh \rho, \\
A^{2x}_\rho &= \varepsilon (1 + c_s^2) \cosh \rho + 2q \sinh \rho, \\
A^{2t}_q &= \sinh \rho, \\
A^{2x}_q &= \cosh \rho.
\end{align*}

\vspace{1em}
\noindent
\textbf{From Eq.~\eqref{eq:momrcons_newv} :}
\begin{align*}
A^{3t}_\varepsilon &= c_s^2 \sinh \rho, \\
A^{3x}_\varepsilon &= c_s^2 \cosh \rho, \\
A^{3t}_q &= \cosh \rho, \\
A^{3x}_q &= \sinh \rho, \\
A^{3t}_\rho &= \varepsilon (1 + c_s^2) \cosh \rho + 2q \sinh \rho, \\
A^{3x}_\rho &= \varepsilon (1 + c_s^2) \sinh \rho + 2q \cosh \rho.
\end{align*}

\vspace{1em}
\noindent
\textbf{From Eq.~\eqref{eq:heatflow_newv}:}
\begin{align*}
A^{4t}_n &= -\frac{1}{n} \left( \sinh \rho - \frac{5 \lambda q}{8 \varepsilon c_s^2} \cosh \rho \right), \\
A^{4x}_n &= -\frac{1}{n} \left( \cosh \rho - \frac{5 \lambda q}{8 \varepsilon c_s^2} \sinh \rho \right), \\
A^{4t}_\varepsilon &= \frac{\sinh \rho}{\varepsilon} - \frac{10 \lambda q}{8 \varepsilon^2 c_s^2} \cosh \rho, \\
A^{4x}_\varepsilon &= \frac{\cosh \rho}{\varepsilon} - \frac{10 \lambda q}{8 \varepsilon^2 c_s^2} \sinh \rho, \\
A^{4t}_q &= \frac{5 \lambda}{4 \varepsilon c_s^2} \cosh \rho, \\
A^{4x}_q &= \frac{5 \lambda}{4 \varepsilon c_s^2} \sinh \rho, \\
A^{4t}_\rho &= \cosh \rho + \frac{5 \lambda q}{8 \varepsilon c_s^2} \sinh \rho, \\
A^{4x}_\rho &= \sinh \rho + \frac{5 \lambda q}{8 \varepsilon c_s^2} \cosh \rho.
\end{align*}

\twocolumngrid
The source vector \( B^{\mu} \) collects the non-derivative terms from the system of equations. Each component \( B^{\mu} \) corresponding to Eqs.~\eqref{eq:ncons_newv}--\eqref{eq:heatflow_newv} is listed below.

\vspace{1em}
\noindent
\begin{eqnarray}
\nonumber
 B^1 &=& 0,\\
 \nonumber
B^2 &=& 0,\\
\nonumber
B^3 &=& 0,\\
\nonumber
B^4 &=& \frac{3qn}{k \varepsilon}.
\end{eqnarray}

\nocite{*}
%\bibliographystyle{apsrev4-2}

%\bibliography{master_references}% Produces the bibliography via BibTeX.

\end{document}